\documentstyle[aps,prl,epsf,preprint,tighten]{revtex} 
\begin{document} 
\draft
\preprint{June 24,1996} 
\title{Defect Lines in the Ising Model \\ and Boundary
States on Orbifolds} 
\author{Masaki Oshikawa$^1$\cite{oshimail}
and Ian Affleck$^{1,2}$\cite{ianmail} }
\address{Department of Physics$^1$ and Canadian
Institute for Advanced Research$^2$, \\ University of British Columbia,
Vancouver, BC, V6T 1Z1, Canada} 
\maketitle 

\widetext

\begin{abstract} Critical
phenomena in the two-dimensional Ising model with a defect line are
studied using boundary conformal field theory on the $c=1$ orbifold. 
Novel features of the boundary states arising from the orbifold
structure, including continuously varying boundary critical
exponents, are elucidated. New features of the Ising defect problem
are obtained including a novel universality class of defect lines and
the universal boundary to bulk crossover of the
spin correlation function.
\end{abstract}

\narrowtext

Boundary conformal field theory\cite{Cardy:fusion} is of considerable
current interest in string theory,\cite{Polchinski} classical
two-dimensional critical phenomena and quantum impurity
problems.\cite{Affleck:reviewKondo}  In the case of the $c=1$ free
boson, compactified onto a circle or radius $r$,  Dirchlet (D) and
Neumann (N) boundary conditions, in which the field or its dual obey
$\varphi = \varphi_0$ or $\tilde \varphi = \tilde \varphi_0$, have been
well studied.\cite{Callan:crosscaps}  By symmetry, the dimensions of
boundary operators are independent of $\varphi_0$, $\tilde \varphi_0$. 
Here we extend this analysis to the $c=1$
$Z_2$ orbifold\cite{Ginsparg:LesHouches} in which
$\varphi$ and $-\varphi$ are
identified, showing that now the boundary dimensions depend on these
parameters and that special features occur at the orbifold fixed
points, $\varphi_0=0$, $\pi r$, $\tilde \varphi_0=0$, $\pi /2r$.  These
results are used to understand the critical behavior of a defect line
in the Ising model.

We consider a two-dimensional Ising model at its critical temperature,
with a
 (horizontal) defect line. Modifying the vertical couplings across the
defect line, for a square lattice on a torus, yields the classical
Hamiltonian: 
\begin{eqnarray}
  {\cal E} &=& -\sum_{i=1,M} \sum_{j=1,N-1}
           [   J_1 \sigma_{i,j} \sigma_{i+1,j}
              + J_2 \sigma_{i,j} \sigma_{i,j+1} ]
\nonumber \\
&&    - \sum_{i=1,M}
           [   J_1 \sigma_{i,N} \sigma_{i+1,N}
              + \tilde{J} \sigma_{i,N} \sigma_{i,1} ] . 
\label{eq:defectham}
\end{eqnarray}
Various exact results have been obtained on this model.
\cite{Bariev:Ising,AbrahamKoSvrakic}
When the bulk couplings, $J_1$ and $J_2$ are tuned to the critical
point, the defect model exhibits a line of critical points depending on
the details of the defect, i.e. the value of $\tilde{J}$ in Eq.
(\ref{eq:defectham}).  The spin-spin correlation function decays along
the defect line with a critical exponent which varies continuously
along the line of critical points.  Other types of defect lines are
possible, including modification of horizontal couplings, introducing
multi-spin couplings, imposing magnetic fields at the line, etc.  8
other universality classes can be readily obtained by cutting the
system at the defect line and imposing independent boundary conditions,
spin-up, spin-down or free on the 2 sides. 

To apply boundary conformal field theory (CFT) to this problem, we must
``fold'' the system at the defect line,\cite{WongAffleck} (See Fig. 1) 
obtaining 2 Ising variables at each point, coupled only at the
boundary, a special case of the Ashkin-Teller (AT) model.  The
corresponding CFT is hence the $c=1$
$Z_2$~orbifold.\cite{Ginsparg:LesHouches} 
We show that all known universality classes of boundary conditions in
the Ising defect line problem, and some additional ones not previously
known,  correspond to D and N boundary conditions on the orbifold
model.  The universal groundstate degeneracy,
$g$,\cite{Affleck:gtheorem} is calculated for all boundary conditions
and the relative renormalization group stability is discussed. We have
calculated the spin correlation function with the spins at arbitrary
points relative to the defect line and each other  for arbitrary
boundary conditions; we give the result for D boundary conditions here.
Details and further results will be given in a longer
paper.\cite{Oshikawa}

The Hilbert Space for the AT model with periodic boundary conditions
contains two sectors, in which the free boson obeys either periodic or
twisted boundary conditions.\cite{Ginsparg:LesHouches}  Boundary states
are constructed, in general, using both sectors. 
(Related discussions in string theory context are found in~\cite{string}.)
For a periodic boson,
boundary states are constructed using the oscillator creation operators
for right and left movers of momentum $\pm 2\pi n/\beta$,
$a_{Rn}^\dagger$ and $a_{Ln}^\dagger$ where $\beta$ is the length of
the periodic spatial interval.  In addition the zero-mode states,
$|(w,k)\rangle$ are used.  Here $w$ and $k$ are integers which label
winding numbers in the space and time direction respectively.  The left
and right-moving energies of these states are $(2\pi /\beta )x$ and
$(2\pi /\beta )\bar x$ with $x=(rw+k/2r)^2$, $\bar x=(rw-k/2r)^2$. The
Dirichlet boundary state which is an eigenstate of $\varphi  (\sigma )$
with eigenvalue $\varphi_0$ is:\cite{Callan:crosscaps} \begin{equation}
\label{eq:DirBS}
  | D ( \varphi_0) \rangle =
  \frac{1}{\sqrt{2 r}}  \exp{[ - \sum_{n=1}^{\infty} a^{\dagger}_{Ln}
a^{\dagger}_{Rn} ] }
   \sum_{k= - \infty}^{\infty} e^{- i k \varphi_0/r }
   | ( 0, k ) \rangle . \end{equation} Orbifold model boundary states
must be invariant under $\varphi \to -\varphi$; hence it is necessary
to combine the D states with eigenvalue $\varphi_0$ and $-\varphi_0$:
\begin{equation}
  | D_O (\varphi_0) \rangle = \frac{1}{\sqrt{2}}
       [ | D(\varphi_0) \rangle + | D( - \varphi_0) \rangle ] ,
\label{eq:bstateboson} \end{equation} It is useful to consider a torus
of circumferences $\beta$ and
 $2l$ with two D defect lines at diametrically opposite locations,
separated by $l$.  After folding, this corresponds to the finite
cylinder geometry considered in Ref. \cite{Cardy:fusion}.  The partition
function, at inverse temperature $\beta$ can be written in terms of the
boundary states as: \begin{equation} Z_{\varphi_0,\varphi_0'}=\langle
D_O(\varphi_0)| e^{-l H^P_\beta} | D_O(\varphi_0') \rangle 
=Z(\varphi_0-\varphi_0')+Z(\varphi+\varphi_0'),\label{ZDD}\end{equation}
where: \begin{eqnarray} \label{Dir:per} Z(\varphi_0-\varphi_0') &=& 
\langle D(\varphi_0)| e^{-l H^P_\beta} | D(\varphi_0') \rangle
\nonumber \\ &=&{1\over \eta (q)}\sum_{n=-\infty}^\infty
q^{2r^2[n+(\varphi_0-\varphi_0')/2\pi ]^2}, \end{eqnarray} with $\eta
(q)=q^{1/24}\prod_{n=1}^\infty (1-q^n)$ and $q=e^{-\pi \beta /l}$.  
$Z(\varphi_0-\varphi_0')$ is the partition function for a periodic
boson.  Note that all states in the Boltzmann sum have integer
multiplicity and that, in the case $\varphi_0=\varphi_0'$, the
groundstate occurs with multiplicity one.  It is this physical
condition which fixed the normalization of the boundary state in Eq.
(\ref{eq:DirBS}) and (\ref{eq:bstateboson}).  In the case of equal
boundary boundary conditions, the finite-size energies give the
dimensions of boundary operators \cite{Cardy:fusion}.  For a periodic
boson the boundary operator spectrum can be read off from Eq.
(\ref{Dir:per}): $x=2r^2n^2+ \hbox{integer}$. It is independent of
$\varphi_0$, as it must be by symmetry.  However, this is not the case
for the orbifold boson; instead, from Eq. (\ref{ZDD}), the boundary
scaling dimensions vary continuously with $\varphi_0$, corresponding to
a line of fixed points.  

Also note that for the orbifold case when $\varphi_0$ is at one of the
fixed points, $0$ or $\pi r$, the groundstate occurs with multiplicity
two, signalling that this is not a well-defined boundary state.  To
obtain well-defined Dirichlet boundary states in these two cases we
must add a component from the twisted sector.  Imposing the twisted
boundary condition $\varphi (\sigma +\beta)=-\varphi (\sigma )$, leads
to oscillators with half-integer momenta, $a_{n+1/2}$.  There are no
zero-modes in this case and the constant part of $\varphi$ must have
the value $0$ or $\pi r$, corresponding to two different oscillator
groundstates, $|0\rangle_T$ and $|\pi r\rangle_T$.  Thus we can only
construct Dirichlet states in the twisted sector for these values of
$\varphi_0$.  These states are: \begin{equation} |D(\varphi_0)_T\rangle
\equiv \exp [{-\sum_{n=0}^\infty a_{L(n+1/2)}^\dagger
a_{R(n+1/2)}^\dagger}]|\varphi_0\rangle_T, \end{equation} where
$\varphi_0=0$ or $\pi r$ {\it only}.  There are four possible D
boundary states in these cases, which give integer multiplicities when
combined with the other D-states and give unit multiplicity for the
groundstate when combined with themselves.  These are: \begin{equation}
|D_O(\varphi_0) \pm \rangle \equiv 2^{-1/2}|D(\varphi_0)\rangle \pm
2^{-1/4}|D(\varphi_0)_T\rangle, \end{equation} for $\varphi_0=0$, $\pi
r$ {\it only}.

In a similar manner we can construct N-states.  For a periodic boson
these are: 
\begin{equation} \label{eq:NeuBS}
  | N ( \tilde{\varphi}_0) \rangle =
  \sqrt{r}
   \sum_{w= - \infty}^{\infty} e^{- 2 i r w \tilde{\varphi}_0}
   \exp{[ + \sum_{n=1}^{\infty} a^{\dagger}_{Ln} a^{\dagger}_{Rn} ] }
   | ( w, 0 ) \rangle . \end{equation}  Note that in this case we use
the other type of zero-modes which vary in the space-direction, and
that the sign in the exponential for the oscillator factor is
reversed.  $| N ( \tilde{\varphi}_0) \rangle$ is an eigenstate of the
dual field, $\tilde \varphi (\sigma )$, of radius $1/2r$, with
eigenvalue $\tilde{\varphi}_0$.  We may regard $\tilde \varphi$ as an
orbifold variable with fixed points at $0$ and $\pi /2r$.  Thus the
orbifold N states are: 
\begin{equation}  |N_O (\tilde \varphi_0)\rangle = 
\frac{1}{\sqrt{2}} [ | N(\tilde{\varphi}_0 \rangle + | N( -
\tilde{\varphi}_0) \rangle ],   \end{equation} for $0<\tilde
\varphi_0<\pi /2r$.  We can construct 4 special N-states for $\tilde
\varphi_0=0$, $\pi /2r$ by analogy with the D construction, using the
twisted sector.  These are constructed from the linear combinations of
oscillator groundstates $\left( | 0 \rangle_T \pm | \pi r \rangle_T
\right)$, respectively.  

These 2 continuous lines, $0<\varphi_0<\pi r$ and $0<\tilde
\varphi_0<\pi /2r$ plus 8 the discrete points of boundary states obey
the Ishibashi condition $[T (\sigma )-\bar T(\sigma )]|A\rangle = 0$
where $T$ and $\bar T$ are the left and right-moving components of the
energy-momentum tensor.  All partition functions constructed from  any
pair of these boundary states contain only non-negative integer
multiplicities.  The states constructed from the same boundary state at
both boundaries all have unit multipicity for the identity operator. 
We conjecture that for generic values of $r$ this is the most general
set of boundary states satisfying these conditions.  The analogous
statement has been proven for a periodic boson.\cite{Friedan}

We now apply the $r=1$ case of  this general analysis of orbifold
boundary CFT, in to the Ising model with a defect line.  The D-boundary
conditions correspond to the integrable defect line of Eq.
(\ref{eq:defectham}).  By comparing the exact
spectrum\cite{AbrahamKoSvrakic} with that of the CFT we obtain:
\begin{equation}\tan (\varphi_0-\pi /4)={\sinh [(1-\tilde J/J_2)J_2/T]
\over \sinh [(1+\tilde J/J_2)J_2/T]} \ \  (0<\varphi <\pi
),\end{equation} while $\sinh (J_1/T)\sinh (J_2/T)=1$ so that the bulk
Ising model is critical.  As we vary $\tilde J/J_2$ from $-\infty $ to
$\infty$, $\varphi_0$ decreases monotonically from $\pi
/4-\tan^{-1}(\exp (2 J_2/T))$ (between $3\pi /4$ and $\pi$) to $\pi
/4-\tan^{-1}(\exp (-2 J_2/T))$ (between $0$ and $\pi /4$).  Only in the
extreme anisotropic limit $J_2/J_1\to 0$, does $\varphi_0$ approach the
endpoint values $\pi$ and $0$.  

There are additional discrete universality classes of defect lines in
which different boundary conditions are imposed independently on the
two sides of the line.  Cardy has shown\cite{Cardy:fusion} that there
are only three universality classes of boundary conditions for an Ising
model on a semi-infinite plane: spin-up ($\uparrow$), spin-down
($\downarrow$) and free ($f$).  We denote the  boundary state
corresponding to a $\uparrow$ boundary condition on the left of the
defect and a $\downarrow$ on the right as $|\uparrow \downarrow
\rangle$, etc.  The $|ff\rangle$ state simply corresponds to the
Dirchlet state with $\tilde J/J_2=0$, $\varphi_0=\pi /2$.  By direct
comparison of the partition functions, we have verified that the four
boundary states corresponding to $\uparrow$ and $\downarrow$ boundary
conditions are: 
\begin{equation} |\uparrow \uparrow \rangle =
|D_O(0)+\rangle ,\ \  |\downarrow \uparrow \rangle = |D_O(\pi )+\rangle, 
\end{equation}
with the spin-reversed states given by the 
corresponding ``-'' orbifold boundary states. It is interesting to note
that the unphysical boundary state obtained by extrapolating
$\varphi_0 \to 0$ without including the twisted sector can be written:
\begin{equation} \lim_{\varphi_0\to 0} |D_O(\varphi_0)\rangle = 
|D_O(0)+\rangle + |D_O(0)-\rangle = |\uparrow \uparrow \rangle +
|\downarrow \downarrow \rangle .
\end{equation} This limit corresponds
to an infinitely strong coupling across the defect line and also a
limit where the other vertical couplings $J_2/T\to 0$
while the horizontal couplings, $J_1/T\to \infty$.  Thus it seems
reasonable to suppose that the two horizontal chains of spins which are
coupled across the defect line, $(i,N)$ and $(i,1)$ get locked into a
perfectly ferromagnetically aligned state corresponding to $|\uparrow
\uparrow \rangle +|\downarrow \downarrow \rangle$.  Similarly, the
limit $\varphi_0\to \pi$ gives the perfectly aligned antiferromagnetic
state.  

The other four possibilities for imposing different boundary conditions
on the two sides of the defect line, $|f \uparrow \rangle$, etc.
correspond to the four endpoint N boundary conditions in a similar
way.  The nature of the N-line of boundary states is commented on below.

We now turn to the RG stability of these various universality classes of
boundary conditions.  This can be addressed by checking for relevant
boundary operators allowed by symmetry, and also by invoking the
$g$-theorem which states that the groundstate degeneracy, g, always
decreases in flows from less stable to more stable fixed points.  We
first consider the continuous line of D-fixed points, with
$0<\varphi_0<\pi$.  This has the value, $g=1$, along the entire line,
indicating that this is indeed a line of stable fixed points; no flow
can occur along the line.  We can read off the dimensions of all
boundary operators from the partition function of Eq. (\ref{ZDD}).  The
only non-trivial relevant boundary operators (of dimension $x\leq 1$)
have dimensions  $2(\varphi_0/\pi )^2$ and $2(\varphi_0/\pi -1)^2$. 
They can be shown to correspond to $\sigma_1\pm \sigma_N$, the sum and
difference of the Ising spin operators on the two sides of the defect
line.  If we consider a defect line with the Ising $Z_2$ symmetry, then
the  relevant boundary operators are prevented from occuring by
symmetry.  Thus the D-boundary conditions are a stable line of fixed
points which should attract generic defect lines with $Z_2$ symmetry. 
That is, we may consider arbitrary vertical and horizontal couplings
near the defect line (which preserve horizontal translational symmetry)
and always expect to renormalize to a D fixed point with some value of
$\varphi_0$.  On the other hand, if we break the $Z_2$ symmetry, for
example by appling a magnetic field along the defect line, then we
destablize these fixed points.  It is natural to expect that we then
renormalize to one of the endpoint D-boundary conditions corresponding
to independent spin up or down boundary conditions on the two sides of
the defect line.  Taking over the results for a semi-infinite Ising
system,\cite{Cardy:fusion,Affleck:gtheorem} we conclude that these
boundary conditions have $g=1/2$,  and no relevant operators.  They are
the most stable fixed points.  

The general N-states have $g=\sqrt{2}$ and have relevant operators, of
dimension $x=1/2$ which correspond to a product of the Ising spin
operators on the two sides of the defect line.  Thus, they represent
unstable critical points which could only be achieved by imposing
additional symmetries at the defect line.\cite{Oshikawa}
We expect that adding other
couplings would produce a flow away from the N boundary conditions to D
boundary conditions ($g=1$), if the $Z_2$ symmetry is preserved, or to
fixed spin boundary conditions ($g=1/2$) if not.  

Finally, we turn to the  critical spin-spin correlation function for an
infinite system with a defect line of D-type,
for arbitrary strength of the defect.
We introduce a complex
spatial co-ordinate for the folded system, $z=x+iy$, where $x$, $y$ are
the two spatial co-ordinates; the defect is at $y=0$ so $y\geq 0$ after
folding.  (If the Ising model does not have $J_1=J_2$, square symmetry,
it is neccessary to rescale one of the co-ordinates.) The folding
introduces two different spin operators, $\sigma_1$ and $\sigma_2$ at
each point.  This universal bulk to boundary crossover function,
depending in a non-trivial universal way on the cross-ratio: $x\equiv
|(z_1-z_2)/(z_1-\bar z_2)|^2$, can be determined by standard methods in
terms of the boundary state.\cite{Cardy:fusion,CardyLewellen}.   Using
the explicit construction of the AT model conformal blocks by
Zamolodchikov\cite{Zam:ATspin}, and the D-boundary states given above,
we obtain: \begin{equation} \langle \sigma_j (z_1)\sigma_j (z_2)\rangle
=  \left( \frac{1}{4 y_1 y_2 x} \right)^{1/8}
\frac{1}{\vartheta_3(u(x))}
          \vartheta_3(e^{2 i \varphi_0}, \sqrt{u(x)}), 
\label{eq:boundary11}\end{equation}  for two spins on the same side of
the defect line (before folding).  Here $\vartheta_j$ are elliptic
theta functions (as defined in Ref.~\cite{Ginsparg:LesHouches})
and $u(x)$ is defined by
$x =[\vartheta_2(u)/\vartheta_3(u)] ^4$ .
When the two points are on
opposite sides of the defect line before folding we obtain:
\begin{equation} \langle \sigma_1 (z_1)\sigma_2 (z_2)\rangle =  \left(
\frac{1}{4 y_1 y_2 x} \right)^{1/8} \frac{1}{\vartheta_3(u(x))}
          \vartheta_2(e^{2 i \varphi_0}, \sqrt{u(x)}).
\label{eq:boundary12}\end{equation} Using Mathematica, we have plotted,
in Fig. (2), the correlation functions for the two points at unit
distance from the boundary and separated by an arbitrary distance from
each other.  In the short-distance limit, $\langle \sigma_1 \sigma_1
\rangle$ converges to a unique power-law, 1/4, which is independent of
the defect strength. This is the expected bulk limit. In the
large-distance limit, the correlation function is governed by another
exponent, $(2\varphi_0/\pi )^2$, which depends on the defect strength.
This is the boundary limit. Our result interpolates between these two
limits. In the ``short-distance'' limit, $\langle \sigma_1 \sigma_2
\rangle$ converges to a constant which depends on the defect strength.
This actually corresponds to two spins located symmetrically about the
defect line. In general, $\langle \sigma_1 \sigma_2 \rangle$ is smaller
than $\langle \sigma_1 \sigma_1 \rangle$ for the same defect strength
and the same (horizontal) distance, as expected. Nevertheless, they
asymptotically converge to the same power-law function in the
large-distance limit with the constant prefactor for $\tilde J \neq 0$.

We thank F. Lesage for useful information. This work is partly
supported by NSERC. M. O. thanks the Killam memorial fellowship for
financial support. After most of this work was completed, we were
informed that A.W.W. Ludwig has obtained some related results
independently.


\begin{figure}[htbp]
  \begin{center}
    \leavevmode
    \epsfysize=12cm
    \epsfbox{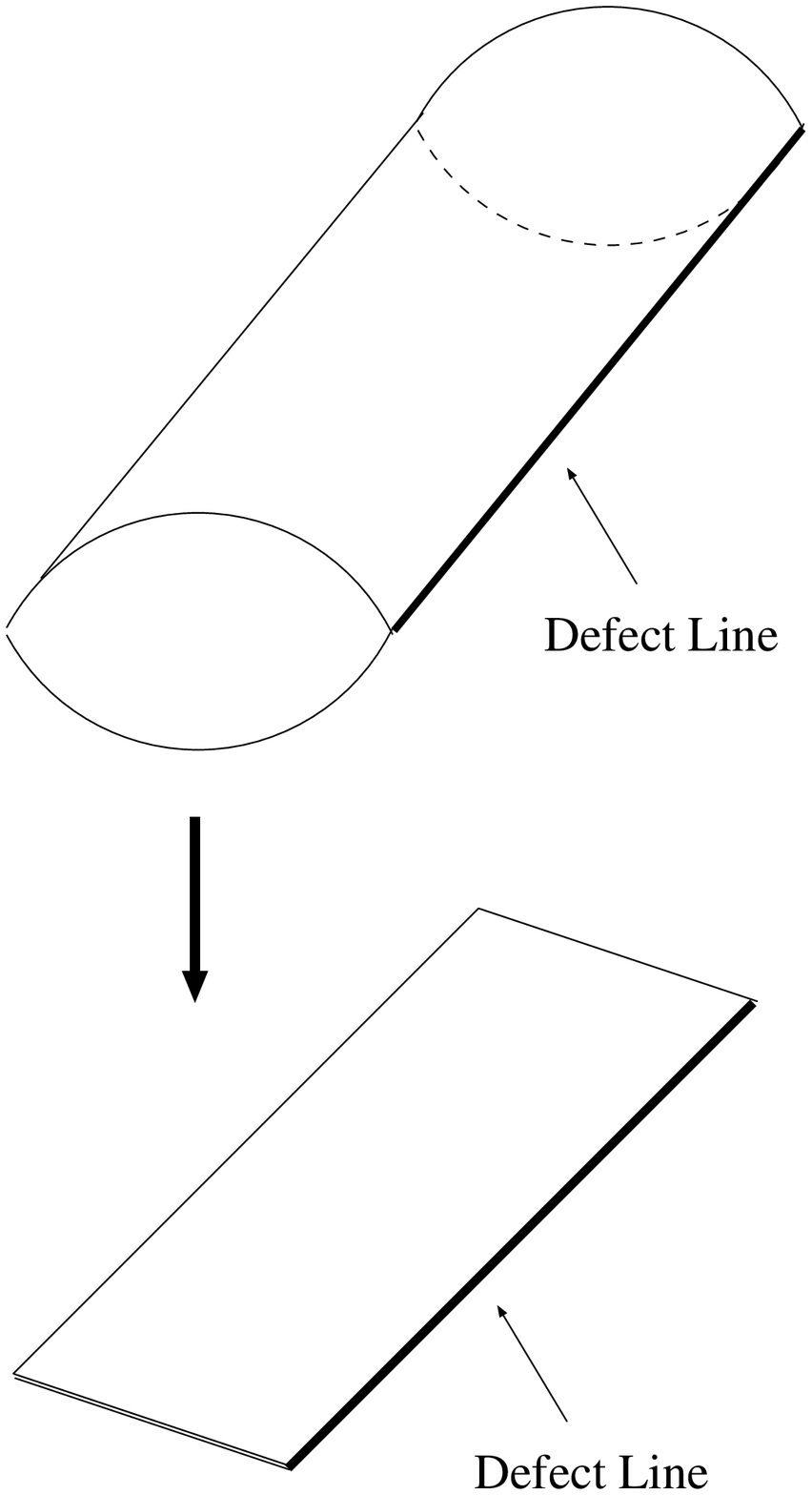}
\caption{The folding of the Ising model on a cylinder to
a $c=1$ theory on a strip. We fold at the defect line and
also at the line on the opposite side.
These lines correspond to the boundary in the folded system.}
\label{fig:fold}
\end{center}
\end{figure}

\begin{figure}
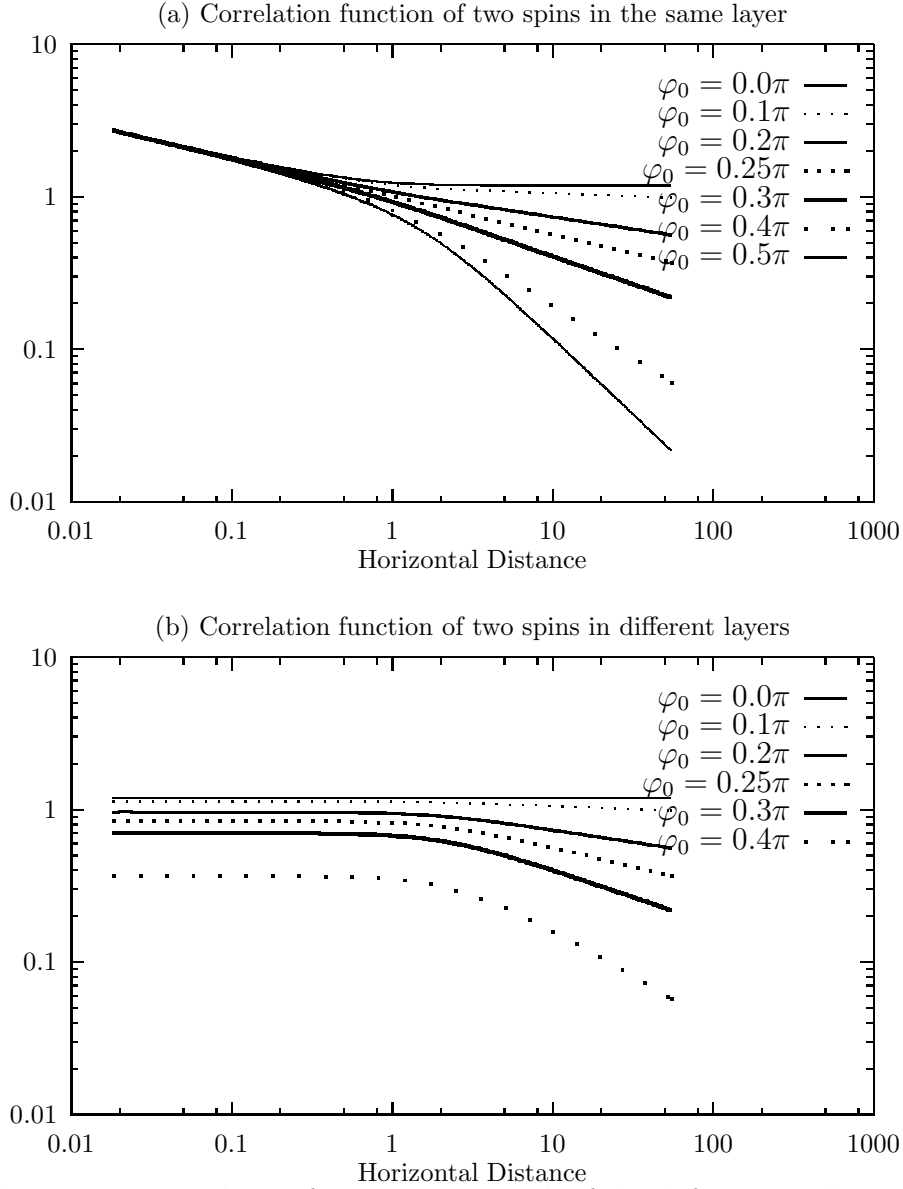

  \begin{center}
    \leavevmode
\input{c11.tex}
    \vspace*{0.5cm}
\input{c12.tex}
\caption{The two-spin correlation for various strength of the defect.
We show the result for (a) $\langle \sigma_1 \sigma_1 \rangle$ and
(b) $\langle \sigma_1 \sigma_2 \rangle$ for
$\varphi_0 = 0$ (strong coupling and anisotropic limit), 
$0.1 \pi , 0.2 \pi , 0.25 \pi$ (no defect)  , $0.3 \pi , 0.4 \pi$
and $0.5 \pi$ (free boundary condition).
They are shown as a function of the
(horizontal) distance $r$, in a log-log plot.}
\label{fig:graph}
\end{center}
\end{figure}

\end{document}